# Calculation technique for simulation of wave and fracture dynamics in a reinforced sheet

*M. Ayzenberg-Stepanenko[a], Z. Yanovitsky[a] and G. Osharovich[b]*

[a]Ben-Gurion University of the Negev, Beer-Sheva, Israel

[b]Bar-Ilan University, Ramat-Gan, Israel

**Abstract.**

Mathematical models and computer algorithms are developed to calculate dynamic stress concentration and fracture wave propagation in a reinforced composite sheet. The composite consists of a regular system alternating extensible fibers and pliable adhesive layers. In computer simulations, we derive difference algorithms preventing or minimizing the spurious distortions caused by the mesh dispersion and obtain precise numerical solutions in the plane fracture problem of a pre-stretched sheet along the fibers. Interactive effects of microscale dynamic deformation and multiple damage in fibers and adhesive are studied. Two engineering models of the composite are considered: the first assumes that the adhesive can be represented by inertialess bonds of constant stiffness, while in the second one the adhesive is described by inertial medium perceived shear stresses. Comparison of results allows the evaluation of facilities of models in wave and fracture patterns analysis.

## 1. Introduction

Since fiber-reinforced sheets have a wide range of applications they may be required to endure intense dynamic loadings ([1-5]). A set of dynamic problems has been studied on the basis of the homogenization approach averaging local properties of microstructure (see e.g. [6]). Within such an approach, features inherent in impact processes are, as a rule, explored by analytical methods and models of solid dynamics that are well developed and intended for analysis of a long-wave spectrum, where microstructure peculiarities play a secondary role. At the same time, the significant rise in influence of microstructure on the wave picture essentially restricts capabilities of analytical modeling in dynamics; gaps of wave fronts and high-gradient components responsible for dynamic stress concentrations and fracture propagation in composite structures have not yet been adequately examined and remain a subject of contemporary research (see e.g. [8 - 10]). These processes are strongly significant, notably in fracture dynamics of reinforced structures where breaking a unit fiber can result, due to wave reflections, in the appearance of multiple wave fronts and high-frequency oscillations.

There exists a set of numerical approaches intended for computer simulations of wave and fracture propagation in homogeneous structures and composites [7-9]. A review of new software upgrades that extend the digital advantage in the composites marketplace can be found in [10]. The main usage of calculation devices based on finite differences, finite elements or spectral elements techniques and their applications to simulation of wave propagation and fracture dynamics in inhomogeneous solids and composite structures are described.

The use of any numerical techniques is advantageous in the case when the specific effects caused by discretization are separated from the expected solution of the original problem, or their influence has been minimized. However, those requirements to preventing numerical errors are not always sufficiently performed in designing of the algorithm. In contemporary literature devoted to the numerical solution of initial-boundary problems, the three main types of numerical errors are considered (see, e.g., [11]): the amplitude error (caused by mesh diffusion and instability), the phase error (mesh dispersion), and the Gibbs error (caused by using a finite representation of a continuous function). To this set the approximation error is to be added inherent to any discretization procedure. If the latter can be somehow estimated in advance, and the stable numerical solution can be justified by the known condition, the numerical diffusion responsible for spreading discontinuities and the numerical dispersion together with Gibbs effects identified as the emergence of spurious oscillations in discontinuity or high-gradient areas remain the main obstacle in preventing sufficient computations on the basis of explicit algorithms. These obstacles turn out be notably significant in simulation of fracture dynamics in reinforced composites: breaking of fibers and cracking of adhesive result in the appearance of multiple wave fronts and high-frequency oscillations. Below the above–mentioned mesh effects are not presented separately (since they have similar reasons and similar consequences) and are all called – Mesh Dispersion (MD).

The purely mathematical aspects related to the MD minimization problem has a half-century history beginning, probably, from the pioneering works [12-14]. Subsequent studies that treat the emergence of MD as a consequence of the Gibbs effect can be found, e.g., in [15 - 18]. In the last decades, a great amount of studies has been devoted to various spatial and temporal discretizations of differential operators to minimize the MD in computation of elastic waves [19, 20], transportation processes [21], and electromagnetic waves [22-24]. The problem of separation of MD oscillations and physical high-frequency oscillations inherent dynamics of heterogeneous solids has not yet been adequately examined and remain a subject of contemporary research (see e.g. [24-26, 36]). Note that calculation results in which spurious oscillations are discussed as effects inherent to the processes appear in some papers even today (see, e.g., [37]).

Among means for preventing MD, we note computational algorithms built on the basis of the so-called Mesh Dispersion Minimization (MDM) technique. The idea behind MDM is to properly adjust the domains of dependence determined by continuous and corresponding discrete models. The MDM algorithms possess the possibility to obtain numerical solutions of transient problems with the same (high) accuracy for low-frequency and high-frequency components. The method initially proposed in [7] was applied to the simulation of fracture dynamics in reinforced fiberglass materials [27, 28]. However in the MDM version used therein, the spurious effects of MD have not been completely eliminated. The MDM technique upgraded in [29] has been used in diverse practical problems: high-speed penetration of metal-fabric composite shields [30], impact indentation of a rigid body into solids [32], and resonant excitation of lattice structures [33]. While in 1D problems MD is completely eliminated by the MDM technique, such a result has not yet been achieved for multidimensional processes.

In this work we present a version of MDM intended for a precise calculation of 2D wave and fracture processes in pre-stretched reinforced composite sheets.

## 2. Mechanical and mathematical models

### 2.1. Physical assumptions

We consider in-plane elastic waves and dynamic fracture in a thin plate modeled by a periodic structure shown in Fig. 1.

The elastic plate of a unit width consists of an unbounded periodic system of high strength fibers, which alternate with pliable adhesive layers. Parameters of fibers: width, $h$, Young modulus, $E$, and density, $\rho_f$; the adhesive has width, $H$, shear modulus, $G$, and density, $\rho_a$. The axis $x$ is directed across the fibers, and the axis $y$ – along them. At infinity ($y = \pm\infty$), fibers are stretched along the axis $y$ by constant tensile stress $\sigma_\infty$.

The mechanical model assumes that fibers function in tension-compression, while the adhesive is under shear stress. The assumption that in fibers exist only normal stresses while in the adhesive only tangential stresses – is often used in studying the static and dynamic equilibrium of unidirectional composites (see, e.g., [34, 35]) having a wide range of practical applications (for example, in aircraft and ship engineering). Although the stress state of structure components is, in fact, more complex, such an approach correctly expresses the concept of the efficient performance of a reinforced material: high strength fibers are oriented along the tensile stress lines, while the adhesive facilitates a more uniform distribution of these loads between fibers, preventing stress concentrations.

The following problem formulation is considered. At time $t<0$ all fibers are stretched along the axis $y$ by a constant tensile stress $\sigma_\infty$ applied at infinity ($x=\pm\infty$). Let one of the fibers (say fiber $m=0$) be suddenly fractured in a cross section (say in $x=0$) at zero moment of time, $t=0$ – see Fig. 1(b). As a result, the initial split increases with time – Fig. 1(c). The problem is to describe the dynamic concentration of stresses, subsequent fracture process occurring with time and to reveal the "trauma" area after the (possible) fracture arrest.

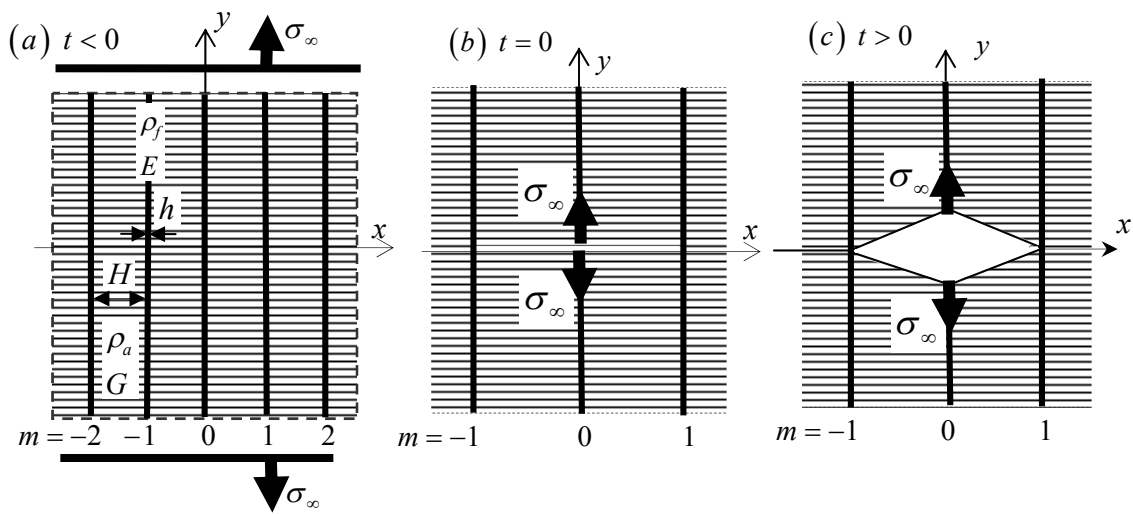

Fig. 1 Schematization of the problem. (a) The plate generated by units consisting of a thin fiber and a thick adhesive layer (their geometrical and physical parameters are shown). The plate is pre-stretched by tensile stresses ($t<0$); (b) At $t=0$ the fiber $m=0$ is suddenly broken in a cross-section $y=0$; (c) Then, at $t>0$, the initial slit, as well as shear stresses at the fiber-adhesive interface increase, unloading the fiber $m=0$ and overloading neighboring fibers $m=1$ and $m=-1$.

We assume $h \ll H$ and introduce a local coordinate $X$ varied in the interval $0<X<H$, thus $x=mH+X$, $(m=0,\pm1,\pm2,\ldots)$. Displacements and stresses in the fibers are denoted by

$u_m(y,t)$ and $\sigma_m(y,t)$, in the adhesive – $v_m(X,y,t)$ and $\tau_m(X,y,t)$. The initial stressed state of the fibers is $\sigma_m = \sigma_\infty$ ($t<0$). It was also assumed that initial tensile stresses in the fibers are less than the given critical value: $\sigma_\infty < \sigma_*$. If $\sigma_m(y,t) > \sigma_*$, the initial fracture can excite a consequence breaking of fibers. In addition, we introduce the critical value of shear stresses in the adhesive, $\tau_*$: if $\tau_m > \tau_*$, the shear crack appears in the corresponding point of adhesive and it can propagate away from the area of the initial fracture.

At $t \geq 0$, the physical pattern of the process can be developed due to the following scenario:

– at $t=0$, two free boundaries appear in the fractured cross-section: $\sigma_0(+0,t) = 0$ and $\sigma_0(-0,t) = 0$. At $t>0$, the unloading wave begins propagating in the fiber, and its free ends move along the axis $y$ in opposite directions;

– along with unloading, the broken fiber pulls the adhesive adjacent to this area. Under the condition that the adhesive does not reveal resistance to normal stresses, the line $y=0$ in adhesive begins to open freely: a transverse crack appears in the adhesive between fibers $m=-1$ and $m=1$, and the splitting area increases with time resulting in the intensification of shear waves in the adhesive, see Fig. 1(b);

– the shear waves reach fibers $m=\pm1$ nearest to the broken fiber $m=0$, overload them (i.e. in addition to $\sigma_\infty$) and involve them in motion. In turn, fibers $m=\pm1$ excite shear waves in the adhesive that propagate upward to the next fibers, and backward to the broken one. Shear waves reflected from the first intact fibers reach the broken one and decelerate it, while the next intact fibers begin to overload. Together with this, free ends in fibers $m=\pm1$ and shear cracks in the nearest adhesive can appear if normal and shear stresses are to exceed critical values $\sigma_*$ and $\tau_*$, respectively. These boundaries surround a domain associated to the trauma despite the fact that in some cases such trauma is not continued but can be formed from a set of alternative fractured and intact domains.

– the dynamic process is developed with time farther on according to the described scheme, more and more fibers neighboring the broken one are being fractured and involved into the motion, while the intact ones get an additional load. At some particular moment of time, energy of the initial fracture is completely spent on tension-compression waves in the fibers and shear waves in the adhesive propagated away of the initial impact area. After that, strains and displacements of the composite reach a new static state, and the dynamic process is finished. As a result, a trauma remains in this new static state.

The used model of the fiber dynamics describes the one-dimensional wave process in a thin rod embedded into the adhesive, while two models are considered for the adhesive:

(i) Model 1. This model is a simplified one. It is designed under the assumption that the adhesive can be represented by inertialess bonds of constant stiffness $K = G/H$. Note that the inertia of bonds cannot be taken into account due to continuous distribution of the bond mass along the fiber in each fiber-adhesive layer.

(ii) Model 2. This model is more precise. The adhesive is described by inertial bonds perceived shear stresses, while tension-compression stresses in bonds are neglected. Such a theoretical treatment of the nature of the components performance is partially justified by the fact that the shear modulus of the adhesive is hundreds of times less than that of the fiber (see e.g. [1-4]), while their stretches have roughly the same level due to the cohesion of the fibers and the adhesive.

In Model 1, the motion of the composite is completely described by displacements of fibers, while displacements of adhesive is calculated from the linear dependence on $X$:
$$v_m(X, y, t) = u_m(y, t) + (X/H)\left[u_{m+1}(y, t) - u_m(y, t)\right].$$

The following patterns can be realized in the considered models depending on the strength of the conditions:

− the composite remains intact: maximal stresses reached in the fibers and the adhesive do not exceed critical limits: $\sigma_m < \sigma_*$, $\tau_m < \tau_*$. These conditions are used to calculate the parameters of stress concentration in the intact fibers and the adhesive associated to them;

− fibers are fractured, while the adhesive remains intact: $\sigma_m \geq \sigma_*$, $\tau_m < \tau_*$;

− adhesive is fractured, while the fibers (with $m \neq 0$) remain intact: $\sigma_m < \sigma_*$, $\tau_m \geq \tau_*$;

− both fibers and adhesive are fractured: $\sigma_m \geq \sigma_*$, $\tau_m \geq \tau_*$.

## 2.2. Mathematical formulation

In the mathematical sense, we are dealing with a non-linear hyperbolic problem possessing non-classical boundary conditions. Due to the symmetry, a quarter of the plane $x, y$ can be considered in the calculation algorithm (let it be $x \geq 0$, $y \geq 0$) with the symmetry

condition in mind. Displacements and strains of the composite at the intact static state ($t < 0$) are

$$u_m(y) = y\sigma_\infty/E, \text{ or } \varepsilon_m(y) = \sigma_\infty/E; \quad v_m(X,y) = 0 \quad (m = 0, \pm 1, \pm 2, \ldots), \tag{1}$$

where $\varepsilon_m(y) = \partial u_m/\partial y$ is the strain in $m^{th}$ fiber ($m = \pm 1, \pm 2, \ldots$), and the fracture event of the fiber $m = 0$ at $t = 0$ changes (1) by adding condition $\varepsilon_0(0,t) = 0$:

$$\varepsilon_0(0) = 0; \quad \varepsilon_m(y) = \sigma_\infty/E \ (m \neq 0, y \neq 0), \quad v_m(X,y) = 0 \ (m = 0, \pm 1, \pm 2, \ldots). \tag{2}$$

Let us reformulate the problem for the additional dynamic state. For this we subtract the static strains (1) from (2). Then boundary conditions for strains in fibers are the following:

$$y = 0: \quad \varepsilon_0(0,t) = -\sigma_\infty/E, \quad \varepsilon_m(0,t) = 0 \ (m \neq 0). \tag{3}$$

The motion of fibers is described by the 1D wave equation

$$\rho_f h \frac{\partial^2 u_m}{\partial t^2} = Eh \frac{\partial^2 u_m}{\partial y^2} + \tau_m^+(y) - \tau_m^-(y), \quad m = 0, \pm 1, \pm 2, \ldots \tag{4}$$

where $\tau_m^+$ and $\tau_m^-$ correspond to reactive shear forces at the fiber-adhesive interface on the right and the left, respectively. Their expressions in Model 1 are:

$$\tau_0^+ = -\tau_0^- = K(u_1 - u_0) \ (m = 0), \quad \tau_m^+ = K(u_{m+1} - u_m), \quad \tau_m^- = K(u_m - u_{m-1}) \ (m > 0), \tag{5}$$

while in Model 2 (recall, here adhesive is considered as a array of inertial bonds), reactive forces are the following:

$$\tau_0^+ = -\tau_0^- = G(\partial v_0/\partial X)\big|_{X=0}, \quad \tau_m^+ = G(\partial v_m/\partial X)\big|_{X=0}, \quad \tau_m^- = G(\partial v_{m-1}/\partial X)\big|_{X=H} \ (m > 0), \tag{6}$$

and displacements in adhesive described by wave equations

$$0 \leq X \leq H: \quad \frac{\partial^2 v_m(X,y,t)}{\partial t^2} = c_a^2 \frac{\partial^2 v_m(X,y,t)}{\partial X^2}, \quad (c_a = \sqrt{G/\rho_a}), \quad m = 0, 1, 2, \ldots \tag{7}$$

with the following boundary conditions:

$$v_m(0,y,t) = u_m(y,t), \quad v_m(H,y,t) = u_{m+1}(y,t). \tag{8}$$

Then we add fracture conditions to the system (3)-(8) as follows. If in the current cross section $\tilde{y}$ of $m^{th}$ fiber at the moment of time $t = t_m^*$ tension stress $\sigma_m(\tilde{y})$ reaches the critical value $\sigma_*$, this cross section breaks, and new boundaries appear resulting in the following conditions for equations (4):

$$\sigma_m(\tilde{y}+0,t) = \sigma_m(\tilde{y}-0,t) = 0 \quad (t > t_m^*). \tag{9}$$

In the case of the fiber-adhesive splitting, we, in addition to (7), have the additional expressions for reactive forces in equations (4) and boundary conditions for $v_m(0,y,t)$ and $v_m(H,y,t)$:

$$\tau_m^\pm\left(\xi^\pm,\tilde{y},t\right) \geq \tau_* \Rightarrow t_{m,\pm}^* = t; \quad t > t_{m,\pm}^*: \quad \tau_m^\pm\left(\xi^\pm,\tilde{y},t\right) = 0, \quad \partial v_m^\pm\left(\xi^\pm,\tilde{y},t\right)/\partial X = 0, \quad (10)$$

where $\xi^+ = 0$, $\xi^- = H$, and indices "$\pm$" at $t_m^*$ denote right and left interfaces, respectively.

It is evident fact that each possible scenario of wave-fracture pattern described above in Subsections 2.1 and 2.2 is saturated by reflected waves and discontinuities that appeared due to fiber rapture snaps and adhesive delaminating. Our goal is to calculate such processes as precise as possible. Below we present the principle and practical devices of the MDM technique allowing this goal to be reached.

## 3. MDM calculation algorithm

To calculate the system (3)-(10), we use the explicit finite difference algorithm. Let temporal and spatial steps of the difference mesh be $\Delta t$, $\Delta x$ and $\Delta y$. Let also values $E, \rho_f, h$ be measurement units, then $c_f = \sqrt{E/\rho_f}$ is the velocity unit.

Calculation algorithms for a discrete analog of system (3)-(10) are constructed on the basis of the mesh dispersion minimization (MDM) technique elaborated in [7] and upgraded in [29]. Simple examples presented below allow the main principle of the MDM to be elucidated.

### 3.1. Free fiber

First, we consider the classic wave problem for a semi-infinite straight elastic fiber subjected to the step tensile stress, which affects the free end. Below we have presented the equation of motion with the step boundary condition at the free end $y = 0$, (11), the plane wave solution for free waves in the fiber and the dispersion relation, (12), the d'Alembert's solution for the stress propagating along the fiber, (13):

$$\ddot{u} = u'' \quad (\dot{u} = \partial/\partial t, \; u' = \partial/\partial y), \; \sigma(0,t) = u'(0,t) = \sigma_0 H(t), \quad (11)$$

$$u(y,t) = U \cdot \exp\left[iq(ct-y)\right] \Rightarrow c = \omega/q = c_f \equiv \sqrt{E/\rho_f} = 1, \quad (12)$$

$$\sigma(y,t) = \sigma_0 H(t-y), \quad (13)$$

where *H(z)* is the Heaviside step function, *U* – const, *q* is the wave number ($q = 2\pi/l$, *l* is wavelength), $i = \sqrt{-1}$, and *c* is the phase velocity (below for simplicity $\sigma_0 = 1$ is taken). We also postulate zero initial conditions.

Dispersion relation $c = 1$ proves the well-known fact – the dispersionless wave propagation in this system: the phase velocity is independent of the wave length. The domain of dependence of $(y,t)$ is bounded by the line (or characteristics) $y = t$.

Discrete analogues of equations (11) and (12) built on the basis of the explicit difference scheme (five-point stencil) are:

$$u_j^{k+1} - 2u_j^k + u_j^{k-1} = \lambda^2 \left( u_{j+1}^k - 2u_j^k + u_{j-1}^k \right), \quad u_{-1}^0 = u_0^0 + \Delta y, \tag{14}$$

$$u_j^k = U \cdot \exp\left[ iq(k\Delta t - j\Delta y) \right] \Rightarrow c = (2/q\Delta t)\arcsin\left[ \lambda \sin(q\Delta y/2) \right], \quad \lambda = \Delta t/\Delta y, \tag{15}$$

where indices *k* and *j* are current numbers of temporal and spatial steps within the mesh $y = j\Delta y$, $t = k\Delta t$; $j,k = 0,1,2,\ldots$, so that $u(y,t) \Rightarrow u(j\Delta y, k\Delta t)$, $\lambda$ is the Courant number. Below $\Delta y$ is taken as the length unit: $\Delta y = 1$, then $\lambda = \Delta t$.

One can see from (15) that phase velocity $c = c(q, \Delta t, \Delta y)$ is not constant in the discrete model – waves possess the dispersion caused by the discretization: parameters of length and time are appeared in equations of discrete problem (14)-(15) contrary to the continuous problem (11)-(13) of no parameters. This phenomenon is called the mesh dispersion that has no physical nature. Besides, due to the periodicity of *arcsin*, the characteristic equation (15) possesses infinite number of modes (contrary to the continuous case with the single mode). Note that in the longwave spectrum ($l \to \infty$, $q \to 0$), the asymptote of the phase velocity coincides with that obtained in the continuous model. With decrease in *l*, the divergence in velocities rises and reaches maximum at $l = 2$ ($q = \pi$), i.e. at the minimal wavelength described by the difference model.

It turns out, however, that if we set $\lambda = 1$ (the limiting value of the Courant stability criterion: $\lambda \leq 1$) characteristic equation (15) is the same as that in the continuum model: $c = 1$, and domains of dependence coincide (we mean the first mode of the discrete model).

Dispersion properties of (15) are illustrated in Fig.2 (*a*): the straight line, *c* = 1, is related to the continual problem (11)-(13), as well as to the discrete one (14)-(15) at $\lambda = 1$. For $\lambda > 1$ (phase velocities for $q > q_*$ are complex), the difference problem becomes unstable. The dispersion curve for $\lambda = 0.01$ is practically the same as that in the case of the simple spatial discrete (but temporally continuous) model of a linear mass-spring chain (below, the chain): inertial particles linked by inertialess springs. If natural parameters of such a system are taken as measurement units, its characteristic equation is the following: $c(q) = 2\sin(q/2)/q$, $c(\pi) = 2/\pi$.

Note, that we did not succeed in obtaining a closed analytical solution of Eqn. (14) for an arbitrary $\lambda$. However, in the 'dispersionless' case $\lambda = 1$, Eqn. (14) simplifies to the following: $u_j^{k+1} = u_{j+1}^k + u_{j-1}^k - u_j^{k-1}$, $u_{-1}^0 = u_0^0 + 1$. Its solution is obtained by the mathematical induction is:

$$\sigma_i^k = H(k-i) \tag{16}$$

which coincides with the d'Alembert's solution (13) in mesh nodes and integer values of time (recall that $c_f = 1$, $\sigma_0 = 1$ and $\Delta y = 1$ are taken, the difference $k - j$ is the discrete analog of the characteristics $y - c_f t$). Similarly to this, any loading form in problem (11)-(13) will be accurately described as the d'Alembert's solution by the difference scheme (14)-(15) with $\lambda = 1$ in integer points along the $x$ axis.

In Fig.2 (b) and (c), analytical and computer solutions are compared in mesh nodes at different loadings. If $\lambda = 1$ they coincide, otherwise computer solutions possess pronounced spurious oscillations caused by the MD.

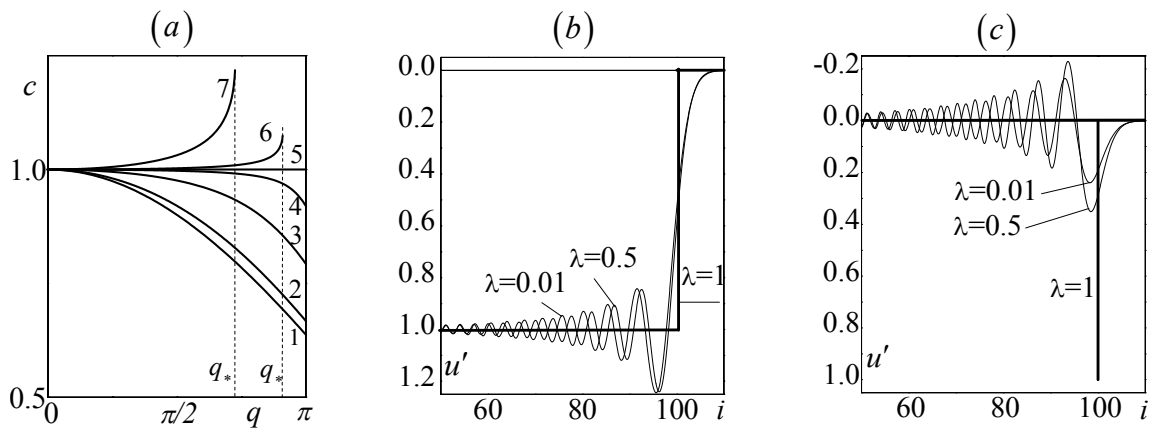

Fig. 2 Free waves and unsteady processes in problems (11) and (12):
(a) – dispersion curves (numbers 1, 2, 3, 4, 5, 6 and 7 correspond to $\lambda = 0.1$, 0.5, 0.9, 0.99, 1.00, 1.01 and 1.05), phase velocities become complex at $q > q_*$; (b) and (c) – stress $\sigma$ versus $x$ (or $j$) at $t = 100$: (b) – the Heaviside step loading, $\sigma(0,t) = H(t)$, (c) – the Dirac delta loading, $\sigma(0,t) = \delta(t)$.

In the latter example, these oscillations do not allow to approach the Dirac delta. Using solutions for wave propagation in the chain [33], the following asymptotes ($\lambda \to 0$ and

$t(k) \to \infty$) are proved for the two mentioned types of loading:

$$\mathbf{1.}\ \sigma(0,t) = H(t) \Rightarrow \sigma(y,t) \sim \left[\frac{1}{3} - \int_0^\eta Ai(z)dz\right],$$

$$\mathbf{2.}\ \sigma(0,t) = \delta(t) \Rightarrow \sigma(y,t) \sim \frac{1}{(k\lambda/3)^{1/3}} Ai(\eta),$$
(17)

where $\eta = \dfrac{k\lambda - i}{(k\lambda/3)^{1/3}}$ and $Ai(z) = \dfrac{1}{\pi}\int_0^\infty \cos(z\tau + \tau^3)d\tau$ is the Airy function. A comparison shows that the maximal divergence between computer solutions depicted in Fig. 2 (*b*) and (*c*) at $\lambda = 0.01$ do not differ from asymptotes (17) more than the curve thickness. Asymptote *1* in (17) describes the propagating wave in which a quasi-front of constant magnitude equal to 1/3 moves (instead of the step with magnitude equal to 1), high-frequency oscillations move behind it with decreasing magnitude and increasing frequency. Asymptote *2* proves a decreasing solution with time (as $t^{-1/3}$) instead of the constant pulse. The wave package spreads with time as $t^{1/3}$. All these peculiarities of the numerical solution disappear at the MDM equality, $\lambda = 1$, eliminating spurious MD effects.

### 3.2. Single fiber inside the adhesive layer

The second example, which has been directly related to the considered dynamic problem, is calculation of wave propagation process in a single semi-infinite fiber upon an elastic foundation: the problem formulation presented above is reduced to the motion of $0^{th}$ fiber, while neighboring fibers are at rest: $u_m(y,t) = 0$, $m \neq 0$. To describe the dynamics of this system, we use Eqns. (4) and (5) of the Model 2. It is enough to add to (11) the 'elastic foundation' with stiffness $2K$ to receive the following equation of motion of the considered system together with the boundary condition:

$$\ddot{u} = u'' + gu,\ \sigma(0,t) = u'(0,t) = H(t);\ (u(y,t) \equiv u_0(y,t)),\ g = 2K.$$
(18)

Substituting $t = \bar{t}\cdot g^{-1/2}$, $y = \bar{y}\cdot g^{-1/2}$ into the differential equation (18), we rewrite it as follows (below dashes are omitted):

$$\ddot{u} = u'' + u,$$
(19)

and using the standard plane wave solution obtain the dispersion relation

$$c = \sqrt{1 + 1/q^2}.$$
(20)

It determines the physical dispersion of free waves propagated along the system. The domain of dependence for Eqn. (19) remains the same as that in (11) – $x = t$, the wave front is formed as a sum of short-length waves at $q \to \infty$ ($l \to 0$).

The aim of the MDM procedure is to build a finite difference scheme possessing a dispersion relation that is maximally close to (20) and results in the coincidence of domains of dependence. As above, in the discrete model we use the explicit scheme of the cross type. The conventional difference analogue of (17) is the following:

$$u_j^{k+1} - 2u_j^k + u_j^{k-1} = \lambda^2 \left( u_{j+1}^k - 2u_j^k + u_{j-1}^k \right) + (\Delta t)^2 u_j^k, \quad \lambda = \Delta y / \Delta t, \tag{21}$$

while the dispersion equation is

$$c = \frac{2}{q \Delta t} \arcsin \left( \lambda \sqrt{\sin^2 \frac{q \lambda}{2} + \frac{(\Delta y)^2}{4}} \right). \tag{22}$$

The latter postulates the stability condition of (21) – the existing real values only of the phase velocity – as

$$\Delta t \leq \Delta y \Big/ \sqrt{1 + (\Delta y)^2 / 4}, \tag{23}$$

which is in contradiction with the MDM requirement to coincidence of dependence domains, $\lambda = 1$. To overcome this contradiction, we use the following three-point-approximation (TPA) (proposed in [27]) of the non differential term in (19):

$$u \sim (1/4) \left( u_{l+1}^k + 2u_l^k + u_{l-1}^k \right). \tag{24}$$

Then the discrete equation (21) has the following form

$$u_j^{k+1} - 2u_j^k + u_j^{k-1} = \lambda^2 \left( u_{j+1}^k - 2u_j^k + u_{j-1}^k \right) + \lambda^2 \left( u_{j+1}^k + 2u_j^k + u_{j-1}^k \right) \Big/ 4, \tag{25}$$

It is easy to see that the order of difference approximation for equations (21) and (25) is the same $\sim (\Delta t)^2 + (\Delta y)^2$, while the dispersion relation for (21) is the following:

$$c = \frac{2}{q \Delta t} \arcsin \left( \lambda \sqrt{\sin^2 \frac{q}{2} + \frac{(\Delta y)^2}{4} \cos^2 \frac{q}{2}} \right). \tag{26}$$

In contrast to the conventional approximation $u \sim u_j^k$ resulting in the stability condition (23), in the TPA case, (24), the stability condition remains the same as in the above-mentioned dispersionless model (11): $\lambda \leq 1$, and the MDM requirement $\lambda = 1$ results in the dispersion relation

$$c = \frac{2}{q} \arcsin\left(\sqrt{\sin^2 \frac{q}{2} + \frac{1}{4}\cos^2 \frac{q}{2}}\right). \tag{27}$$

From the entire interval $0 < \lambda \leq 1$ in relation (22), the phase velocity (27) corresponding to $\lambda = 1$ is maximally close to the that received in continual case (20) at entire discrete spectrum, $0 \leq q \leq \pi$. Along with this, waves of the shortest length, allowed by the discrete model, ($q = \pi$), propagates with phase velocity $c = 1$, which corresponds to infinitely short waves in the continuous model: $q \to \infty$ $(\lambda \to 0)$. So domains of dependence of continual, (19), and discrete, (25), models coincide if setting $\lambda = 1$.

Note that because of the inconsistency of spectral media in the two considered models, MD is not completely prevented, but, as it is shown below, its effect becomes minimal in its main part, which manifests itself in the front zone that is eliminated.

In Fig 3 (a), dispersion curves at various $\lambda$ are depicted. Here $\lambda_C \simeq 0.894427$ is the Courant number: the stability lost occurs at $\lambda > \lambda_C$ and $q > q_*$. Bold curves – the analytical solution (20) and MDM solution (27) at $\lambda = 1$, thin curves – the solution of conventional equation (22) at $\lambda = \lambda_C$. The curves corresponding to the continual and MDM solutions (at $\lambda = 1$) are maximally close to each other.

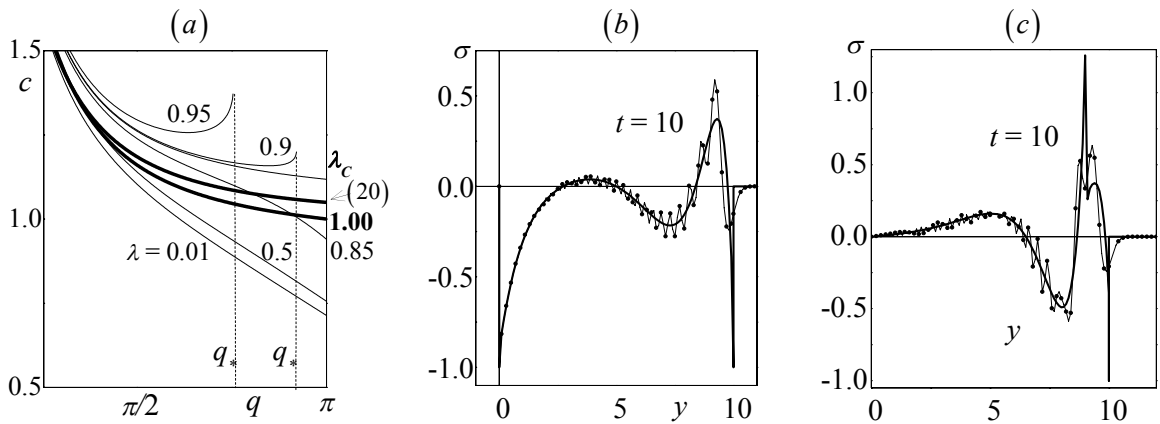

Fig. 3 Free waves and unsteady processes in a fiber upon an elastic foundation. (a) – dispersion curves for several models and $\lambda$: bold curves – the analytical solution of (20) and MDM solution of (27) at $\lambda = 1$, thin curves – solutions of equation (22) for several $\lambda$ ($\lambda_C \simeq 0.894427$ is the Courant number). (b) and (c) – stress $\sigma$ versus $y$ (or $j$) at $t = 10$: (b) – the Heaviside step loading, $\sigma(0,t) = H(t)$, (c) – the Dirac delta loading, $\sigma(0,t) = \delta(t)$.

In Fig. 3(b) and (c), unsteady stress distributions in the fiber along the axis y (at $t = 10$) are depicted ($\Delta y = 0.01$). The loading corresponds to (b) is the Heaviside unit step, $\sigma(0,t) = H(t)$, and (c) – pulse, $\sigma(0,t) = H(10-t)$. Calculation results obtained with the conventional discrete model (21) and $\lambda = \lambda_c$ are shown by thin curves, while bold curves correspond to analytical solutions near the front that can be found in [33] and coinciding with it the MDM-solutions of (25) with $\lambda = 1$. In the case of the Heaviside loading, the above-mentioned analytical solution is as follows:

$$\sigma(y,t) = J_0\left(\sqrt{(t^2 - y^2)}\right) H(t - y), \quad t \sim y, \tag{28}$$

where $J_0(z)$ is the Bessel function of the first kind and zero order.

Comparing calculation results obtained for unsteady problems, we conclude that the MDM solution of (25) with the TPA and $\lambda = 1$ coincide with the analytical representation at the front zone. Probably, this solution plays the role of the analytical one in discrete spatial and temporal coordinates. As to computer results obtained with the conventional scheme (21), the main conclusion is that this model is not suitable to calculate wave processes with fronts and high-frequency vibrations. The corresponding results obtained at $\lambda > \lambda_C$, that determines the best approximation to continual case, prove this assertion.

The results of computer simulations of wave and fracture propagation in the considered fiber-reinforced composite sheet which we have obtained with the use of MDM-algorithms are presented below.

## 4. Results of computer simulations

In Model 1, the MDM calculation algorithm is proved in the case when $c_f \Delta t = \Delta x$ in (4) and the shear deformations in adhesive, (5), is approximated by the following TPA model:

$$u_{m+1} - u_m \approx (1/4)\left[\left(u_{i+1}^k + 2u_i^k + u_{i-1}^k\right)_{m+1} - \left(u_{i+1}^k + 2u_i^k + u_{i-1}^k\right)_m\right]. \tag{29}$$

As to Model 2, the following approximation results in the MDM algorithm ($\Delta t = 1$) are obtained (using Eqns. (4) and (7)):

$$\begin{aligned}
&\Delta y = c_f, \quad \Delta x = c_a \quad (y = j\Delta y, \, x = i\Delta x); \\
&v_{m,j,i}^{k+1} = v_{m,j,i+1}^{k+1} + v_{m,j,i-1}^k - v_{m,j,i}^{k-1}; \\
&u_{m,j}^{k+1} = u_{m,j+1}^k + u_{m,j-1}^k - u_{m,j}^{k-1} + 2f\left(v_{m,j,1}^k + v_{m-1,j,H/\Delta x}^k\right), \\
&f = GH/(Eh\mu), \quad \mu = h\rho_f \Delta y + H\rho_a \Delta x.
\end{aligned} \tag{30}$$

Some results of computer simulations conducted on the basis of MDM algorithms (29) and (30) are presented below. Parameters of the fiber are measurement units: $h = 1$, $E = 1$,

$\rho_f = 1$ ($c_f = 1$); adhesive parameters are $H = 5$, $G = 0.025$, $\rho_a = 0.4$. We set also $\sigma_\infty = 1$, and $\Delta t = \Delta y = 1$, $\Delta x = c_a$ within the MDM algorithm.

### 4.1. Model I

Here we discuss Model 1. In Fig. 4, stresses vs. time are depicted for the case in which the composite remains intact at $t > 0$: $\sigma^*$ and $\tau^*$ are less than corresponding maximal values reached after the initial fracture at $t = 0$. They are observed in the cross-section $y = 0$: $(\sigma_1)_{max} = 1.482$, $(\tau_0)_{max} = 0.073$. Note, if $t > 25 H/c_f$, stresses are close enough to their static limits: $(\sigma_1)_{st} = 4/3$, $(\tau_0)_{st} = 0.053$, which are analytically obtained in [34]. Below, in calculation of fracture problems, the obtained intervals between maximal and static values are used in order to set $\sigma^*$ and $\tau^*$.

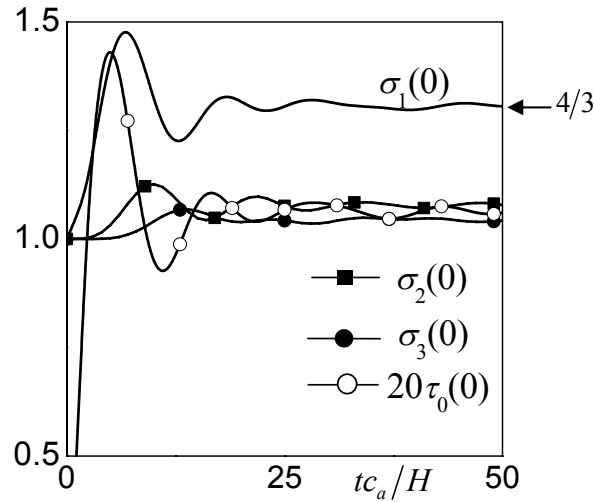

Fig. 4 Model 1. Stresses vs. time in intact composite at $t > 0$

Fracture development with time is shown in Fig. 5. One can observe processes of propagation and arrest of normal cracks – (*a*) and (*b*), and shear cracks – (*c*).

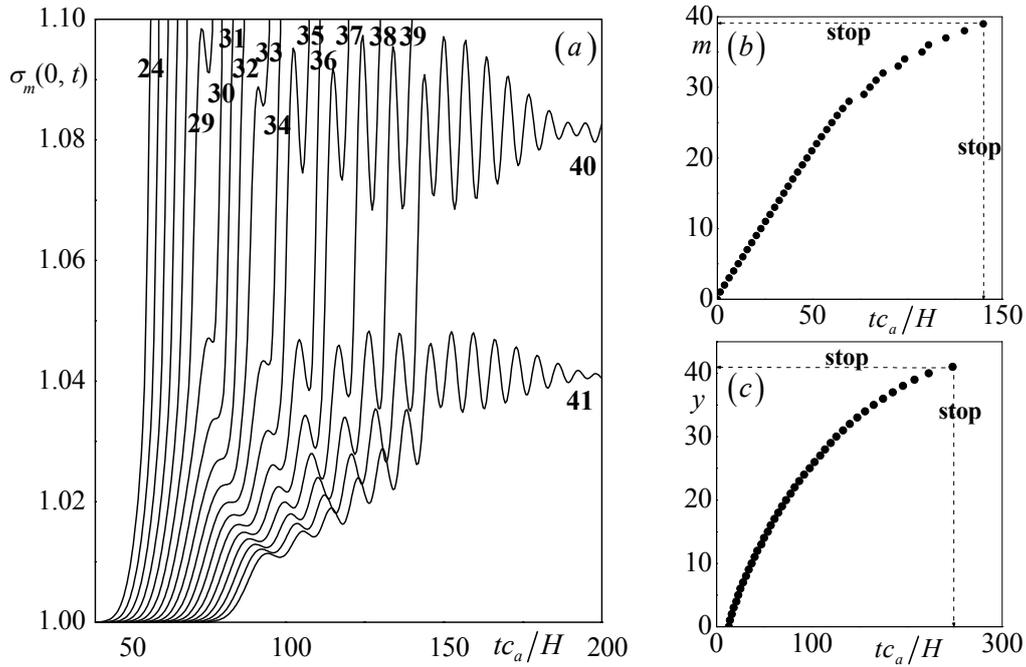

Fig. 5 Model 1. (*a*) and (*b*) – fracture of fibers, $\sigma^* = 1.1$, $\tau^* > \tau_{max}$; (*c*) – fracture of adhesive, $\sigma^* > \sigma_{max}$, $\tau^* = 0.055$. Stresses vs. time in fibers are shown beginning with $m = 24$, fibers with $m > 39$ remain intact (fiber numbers are bold); (*b*) – propagation and arrest of the normal crack in fibers; (*c*) – the same for the shear crack in adhesive.

### 4.2. Model II

Together with results corresponding to wave and fracture dynamics in the considered structure having an independent significance, we use them to identify the scope of the applicability of Model I.

The pattern shown in Fig. 6 is calculated in the case when composite remains intact at $t > 0$. The gap of shear stresses, initially equal to 0.1, doubled after the first reflection from intact fiber $m = 1$ and remains equal to 0.2 in the whole process. After next reflections, duration of the gap decreases, and influence of gaps to normal stresses decreases, as well. As comparison shows, the maximal and static stresses in fibers are practically the same as those obtained for Model 1. But the significant difference in maximal shear stresses (~ 3 times) will result in different fracture processes.

Thus, Model 1 can be used at the initial, pre-fracture stage, where a stress concentration pattern is calculated, while fracture propagation processes saturated by front gaps are to be computed with Model 2. The latter could be used, for example, within computer simulators allowing optimization problems to be explored.

Let, for example, limit $\sigma^*$ be constant, while limit $\tau^*$ is varied within the interval $(\tau_{st}, \tau_{max})$. Our aim is to find such a value of $\tau^*$, for which the volume of breaking fibers, $m^*$ (or, that is the same, the normal crack length equal to $2m^*H$), is less than the given value $M$.

In the table below, we present some related results obtained in the case $\sigma^* = 1.2$ and $M = 10$; in the two lower rows we set the amount (its half due to symmetry) of fractured fibers, $m^*$, and lengths of shear cracks in adhesive, $y^*/h$. One can see that the required result is $\tau^* = 0.1$. Note that in the presented example, shear cracks are observed only in the layer $m = 0$.

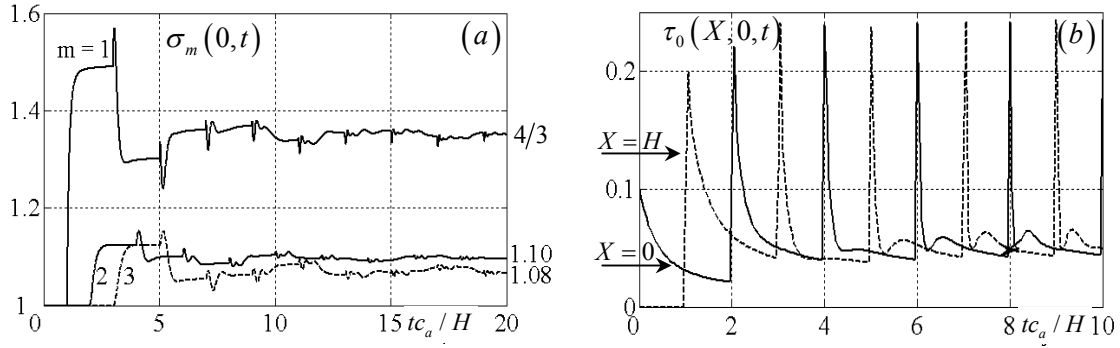

Fig. 6 Model 2 (intact composite at $t > 0$); (a) – normal stresses in fibers, (b) – shear stresses in adhesive (layer $m = 0$) at interfaces $X = 0$ and $X = H$.

Table 1 Simulation results for $\sigma^* = 1.2$ and $M = 10$

| $\tau^*$ | 0.07 | 0.075 | 0.08 | 0.085 | 0.1 | 0.15 | 0.25 |
|---|---|---|---|---|---|---|---|
| $m^*$ | 0 | 0 | 5 | 7 | 9 | 13 | 13 |
| $y^*/h$ | ∞ | 208 | 65 | 23 | 11 | 3 | 0 |

The features of fracture propagation in the case $\sigma^* = 1.2$ and $\tau^* = 0.1$ are shown in Fig.7. After the initial rupture of $0^{th}$ fiber at $t = 0$, the fracture in the adhesive occurring at the fiber-adhesive interface possesses a high-speed avalanche-like pattern; the speed of fracture propagation decreases with time and the adhesive fracture is stopped at $t = 30 H/c_a$. Adhesive layers with $m \neq 0$ remain intact. Fracture of fibers occurs at $y = 0$ and propagates with a practically constant speed up to $t = 63 H/c_a$, then the fracture is arrested.

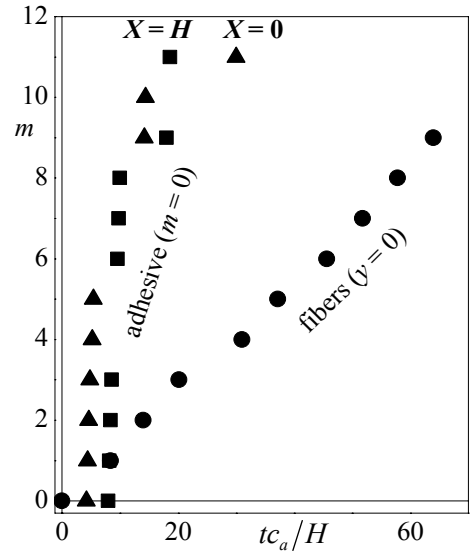

Fig. 7 Model 2: Fracture of fibers and adhesive vs. time ($\sigma^* = 1.2$, $\tau^* = 0.1$)

## 5. Conclusions

Mathematical models and calculation algorithms have been developed to calculate dynamic stress concentration and fracture wave propagation in a reinforced composite sheet. The following aspects have been discussed and analyzed:

- The finite-differences algorithms preventing or minimizing the spurious mesh dispersion and obtaining precise numerical solutions of brittle fracture processes in a pre-stretched sheet along the fibers.

- Interactive effects of microscale dynamic deformation and multiple damage in fibers and adhesive are studied.

- The simulation of dynamic stress concentration and fracture propagation processes in two engineering models of the composite: in the first one the adhesive is represented by inertialess bonds of constant stiffness, while in the second one the adhesive is described by inertial medium perceived shear stresses.

The applicability of simplified models to description of dynamic processes in the composite is discussed.

**Acknowledgments**

This research was supported by The Israel Science Foundation, Grant 504/08.

# References

1. Gibson R F, Wilson D G: 'Dynamic mechanical properties of fiber-reinforced composite materials'. Shock and Vibration Digest 1979 11(10) 3-11.
2. Sela N, Ishai O: 'Interlaminar fracture toughness and toughening of laminated composite materials: a review'. Composites 1989 20(5) 423-35.
3. Summerscales J (ed): Microstructural characterisation of fibre-reinforced composites. Woodhead Publishing, Cambridge, 1998. ISBN 1-85573-240-8.
4. Deng S, Ye L, Mai Y: 'Influence of fibre cross-sectional aspect ratio on mechanical properties of glass fibre/epoxy composites. II. Interlaminar fracture and impact behaviour'. Composites Science & Technology 1999 59 1725-34.
5. Achenbah J D: Vibrations and waves in directional composites. In: Mechanics of Composite Materials 2. N-Y. Academic Press, 1975.
6. Achenbach J D, Sun C T, Herrman G: 'Continuum theory for a laminated medium'. J. Appl. Mech. 1968 35 467–75.
7. Stepanenko M V: 'A method of calculating non-stationary pulsed deformation processes in elastic constructions'. J. Mining Sci. 1976 2 53-57.
8. Langlie S, Cheng W, Dilber I: 'Computer simulation of high-speed impact response of composites'. Proc. 1990 ACM/IEEE Conf. on Supercomputing, N-Y, 1990.
9. Reddy J N, Robbins D H Jr: 'Theories and computational models for composite laminates'. Appl. Mech. Rev. Trans. ASME 1994 47 147-69.
10. McConnell V P: 'Software delivers faster time-to-part and reduced testing'. Reinforced Plastics 2007 51(8) 24-8.
11. Oran E S, Boris J P: Numerical simulation of reactive flow 2$^{nd}$ ed. Cambridge University Press, 2001.
12. Weinberger H F: 'Upper and lower bounds for eigenvalues by finite difference methods' Comm. Pure Appl. Math. 1956 9 613-23.
13. Weinberger H F: 'Lower bounds for higher eigenvalues'. Pacific J. Math. 1958 8 339-68.
14. Lax P D, Wendroff B: 'Difference schemes for hyperbolic equations with high order of accuracy'. Comm. Pure Appl. Math.1964 17 381-98.
15. Fromm J E: 'A method for reducing dispersion in convective difference schemes'. J. Computational Phys. 1968 3 176-89.
16. Serdjukova S: 'The oscillations that arise in numerical calculations of the discontinuous solutions of differential equations'. Comput. Math. and Math Phys. 1971 11 140-54.
17. Chin R C Y: 'Dispersion and Gibbs phenomenon associated with difference approximations to initial boundary-value problems for hyperbolic equations'. J. Computational Phys. 1975 18 233-47.
18. Makarenko A S, and Moskal'kov M N: 'Accuracy and dispersion of difference schemes'. USSR Comp. Math. and Math. Physics 1983 23 4 145-48.
19. Holberg O: 'Computational aspects of the choice of operator and sampling interval for numerical differentiation in large-scale simulation of wave phenomena'. Geophysical Prospecting 1987 35 6 629-55.
20. Jiang L, Rogers R J: 'Effects of spatial discretization on dispersion and spurious

oscillations in elastic wave propagation'. Int. J. Numer. Meth. Engrng 1990 29 1205-18.
21. Fox-Rabinovitz MS: 'Computational dispersion properties of vertically staggered grids for atmospheric models'. Monthly Weather Review 1994 122 2 377-92.
22. Moss C D, Teixeira F L, Jin Au K: 'Analysis and compensation of numerical dispersion in the FDTD method for layered, anisotropic media'. Antennas and Propagation, IEEE Transactions 2002 50 9 1174-84.
23. Oğuz U, Gürel L: 'Reducing the dispersion errors of the Finite-Difference Time-Domain Method for multifrequency plane-wave excitations'. Electromagnetics 2003 23 539-52.
24. Wu Y-S, Forsyth P A: 'Efficient schemes for reducing numerical dispersion in modeling multiphase transport through heterogeneous geological media'. Vadose Zone 2008 J 7 1 340-49.
25. Seriani G, Oliveira S P: 'Dispersion analysis of spectral element methods for elastic wave propagation'. Wave Motion 2008 45 6 729-44.
26. Ogurtsov S, Georgakopoulos S V: 'FDTD schemes with minimal numerical dispersion'. IEEE Trans. Adv. Pack. 2009 32 199.
27. Stepanenko M V: 'Dynamics of the fracture of unidirectional fiberglass'. J. Appl. Math. Tech. Phys. 1979 4 155-63.
28. Stepanenko M V: 'A numerical experiment of the fracture dynamics of a composite material'. Mechanics of. Composite Materials 1981 17 1 46-51.
29. Abdukadyrov S, Alexandrova N, Stepanenko M: 'On MDM approach to calculation of solid and structure dynamics'. J. Mining Sci. 1984 6 34-40.
30. Slepyan L, Ayzenberg-Stepanenko M: 'Penetration of metal-fabric composite targets by small projectiles' Personal Armour Systems. British Crow Copyright/MOD, 1998, 289-98.
31. Kubenko V D, Ayzenberg-Stepanenko M V: 'Impact indentation of a rigid body into elastic layer. Analytical and numerical approaches'. Math. Methods and Phys. Mech. Fields 2008 51 61-74.
32. Ayzenberg-Stepanenko M, Slepyan L: 'Resonant-frequency primitive waveforms and star waves in lattices'. Journal of Sound and Vibration 2008 313 812-21.
33. Slepyan L: Transient elastic waves. St-Petersburg, Sudostroenie, 1972 (in Russian).
34. Mikhailov A M: 'Fracture of a unidirectional glass-plastics'. Mech. of Solids 1973, 5.
35. Mikhailov A M: 'Dynamics of a unidirectional glass-plastics'. J. Appl. Math. Tech. Phys. 1974 4 139-45.
36. Finkelstein B, Kastner R: 'Finite difference time domain dispersion reduction schemes'. J. Comput. Phys 2008 221 422-438
37. Tasdemirci A, Hall I W, Gama B A, Guden M, 'Stress wave propagation effects in two- and three-layered composite materials', Journal of Composite Materials 38 (2004), 995-1010.